\documentclass[aps,pra,superscriptaddress,floatfix,showpacs,10pt,two
column]{revtex4}%
\usepackage[dvips]{graphicx}
\usepackage{amssymb}
\usepackage{amsmath}
\usepackage{color}
\usepackage{amsfonts}%
\definecolor{ueblue}{rgb}{0,0,0.2}
\definecolor{hh}{rgb}{1,0.4,0.3}
\begin{document}
\title{Invariant information and complementarity in high-dimensional states}
\author{Wei Song}
\affiliation{Hefei National Laboratory for Physical Sciences at Microscale and Department
of Modern Physics, University of Science and Technology of China, Hefei, Anhui
230026, China}
\author{Zeng-Bing Chen}
\affiliation{Hefei National Laboratory for Physical Sciences at Microscale and Department
of Modern Physics, University of Science and Technology of China, Hefei, Anhui
230026, China}

\pacs{03.67.-a, 03.65.Ud}

\begin{abstract}
Using a generalization of the invariant information introduced by Brukner and
Zeilinger [Phys. Rev. Lett. \textbf{83}, 3354 (1999)] to high-dimensional
systems, we introduce a complementarity relation between the local and
nonlocal information for $d\times d$ systems under the isolated environment,
where $d$ is prime or the power of prime. We also analyze the dynamics of the
local information in the decoherence process.

\end{abstract}
\maketitle

When dealing with classical measurement the Shannon information
\cite{Shannon:1948} is a natural measure of our ignorance about properties of
a system. However, Shannon information is only applicable when measurement
reveals a preexisting property. In sharp contrast to classical measurements, a
quantum measurement does not reveal a preexisting property. For example, if we
want to read out the information encoded in a qubit, we have to project the
state of the qubit onto the measurement basis $\left\{  {\left\vert
0\right\rangle ,\left\vert 1\right\rangle }\right\}  $ which will give us a
bit value of either 0 or 1. The qubit might be in the eigenstate of the
measurement apparatus only in the special case; in general, the value obtained
by the measurement has an element of irreducible randomness. Thus we cannot
get the value of the bit or even hidden property of the system existing before
the measurement is performed. To overcome this difficulty, Brukner and
Zeilinger introduced the operationally invariant information
\cite{Zeilinger:1999} as the measure of local information. The new measure of
information is invariant under the transformation from one complete set of
complementary variables to another. It is also conserved in time if there is
no information exchange between the system and the environment. In this paper,
using a generalization of Brukner and Zeilinger's result to higher dimensional
case, we derive a complementarity relation between the local and nonlocal
information for $d\times d$ systems.

The new measure is obtained by summing over the measurement outcome of a set
of mutually complementary observables (MCO) \cite{Ivanovic:1981,Wootters:1989}%
. Before describing the details of our derivations, let us briefly review the
definition of MCO. In a $d$-dimensional Hilbert space we call two observables
$\mathcal{K}$ and $\mathcal{M}$ mutually complementary if all their
respective, complete, orthonormal eigenvectors fulfill
\begin{equation}
\left\vert {\left\langle {\mathcal{K},k\left\vert {\mathcal{M},m}\right\rangle
}\right.  }\right\vert ^{2}=\frac{1}{d}{\kern1pt}{\kern1pt}\;\ \forall
k,m=1,\ldots,d.\label{eq1}%
\end{equation}

It is known that if $d$ is prime or the power of prime, the number of MCO is
$d+1$
\cite{Ivanovic:1981,Wootters:1989,Bandyopadhyay:2002,Bengtsson:2005,Romero:2005,Wootters:2004}%
. Consider a measurement of an observable $\mathcal{M}=\sum\nolimits_{j=1}%
^{d}{a_{j}\Pi_{j}}$. Each outcome $j$ is detected with probability
$p_{j}=Tr\rho\Pi_{j}$. According to Ref.~\cite{Zeilinger:1999}, we define the
local invariant information for $d$-level system as
\begin{equation}
I=\mathcal{N}\sum\nolimits_{\alpha=1}^{d+1}{\sum\nolimits_{j=1}^{d}{\left(
{p_{\alpha j}-\frac{1}{d}}\right)  }}^{2},\label{eq2}%
\end{equation}
where $\alpha=1,\ldots,d+1$ and $j=1,\ldots,d$ label complementary observables
and their eigenvectors, respectively, and $\mathcal{N}$ is the normalization
factors. The derivation of our results use the fact that
\cite{Ivanovic:1981,Wootters:1989}
\begin{equation}
Tr\left\{  {\Pi_{\alpha j}\Pi_{\beta k}}\right\}  =\delta_{\alpha\beta}%
\delta_{jk}+\frac{1}{p}\left(  {1-\delta_{\alpha\beta}}\right)  .\label{eq3}%
\end{equation}

After some algebra, we get
\begin{equation}
I=\frac{d}{d-1}\log_{2}d\left(  {Tr\rho^{2}-\frac{1}{d}}\right)  ,\label{eq4}%
\end{equation}
where we choose $\log_{2}d$ as the normalization factor to ensure that a
$d$-level pure state carries $\log_{2}d$ bits of information. If $d=2^{k}$,
$I=\frac{2^{k}k}{2^{k}-1}\left(  {Tr\rho^{2}-\frac{1}{2^{k}}}\right)  $ which
corresponds to the local invariant information for composite $k$-qubit
systems. The special choice for $k=1,2$ recovers the results obtained in
Ref.~\cite{Zeilinger:1999}.

Next, we want to use this new measure to establish a complementary relation
between the local and nonlocal information for $d\times d$ systems.
Complementarity is an important concept in quantum theory. Besides the most
well-known complementarity principle introduced by Bohr \cite{Bohr:1928}, many
other kinds of complementarity relation have also been discussed
\cite{Birula:1975,Deutsch:1983,Krauss:1987,Maassen:1988}. In particular for
two-state systems, elegant relations between two complementary observables
have been derived \cite{Mandel:1991,Jaeger:1995,Englert:1996}. Additionally,
Jakob and Bergou \cite{Jakob:2003} have recently derived a complementarity
relation for an arbitrary pure state of two qubits. Subsequently, their result
was generalized to the multi-qubit systems \cite{Tessier:2005,Bai:2007}.
Recently, Cai \textit{et al}. \cite{Cai:2007} also established an elegant
complementarity relation between local and nonlocal information for qubit
systems. Below we show that a complementarity relation also holds for $d\times
d$ systems.

Now we analyze a two-qutrit system to give an example. Suppose the system
initial state is a pure state $\left\vert \psi\right\rangle _{12}$, and
$\rho_{i}$ ($i=1,2$) is the reduced density matrix of each individual qutrit.
The total information contained in the $3\times3$ system is in two forms. One
is local form, which is the information content in each individual qutrit. The
other is nonlocal form, which is entanglement between the two qutrits. If the
$3\times3$ system is isolated, i.e., it starts with an initial pure and then
is subject to unitary transformations only, then the sum of the local and
nonlocal information should be conserved. Here, we adopt the operationally
invariant information derived above to measure local information. In the case
of $3\times3$ pure state $\left\vert \psi\right\rangle _{12}$, the local
information contained in qutrit 1 and 2 is $I_{1}+I_{2}=\sum\nolimits_{i=1}%
^{2}{\frac{3}{2}}\log_{2}3\left(  {Tr\rho_{i}^{2}-\frac{1}{3}}\right)  $. If
being measured by 2-tangle \cite{Coffman:2000}, the pairwise entanglement of
the pure state $\left\vert \psi\right\rangle _{12}$ is $\tau_{12}=C_{12}%
^{2}=2\left(  {1-Tr\rho_{i}^{2}}\right)  $, where $C_{12}$ is the generalized
notion of concurrence for $3\times3$ pure states
\cite{Wootters:1998,Rungta:2001}. Thus we have the following complementarity
relation
\begin{equation}
I_{1}+I_{2}+\left(  {\frac{3}{2}\log_{2}3}\right)  \tau_{12}=2\log
_{2}3.\label{eq5}%
\end{equation}

Equation (\ref{eq5}) suggests that the nonlocal information has close relation
to entanglement, which is a reasonable fact. As every term of $I_{1}$, $I_{2}$
and $\tau_{12}$ is convex, the total expression is. Therefore, $I_{1}%
+I_{2}+\left(  {\frac{3}{2}\log_{2}3}\right)  \tau_{12}<2\log_{2}3$ for a
mixed $3\times3$ state $\rho_{12}$. This is easy to understand from a physical
picture. If the system is not isolated, the $2\log_{2}3$ bit information is
not only contained in the system, but also in its correlations with the
outside environment. A similar result also holds for $d\times d$ systems if we
notice the following fact. For arbitrary $d\times d$ pure state, we have
$\tau_{12}=C_{12}^{2}\equiv2\left(  {1-Tr\rho_{1}^{2}}\right)  =2\left(
{1-Tr\rho_{2}^{2}}\right)  $. Thus, for a $d\times d$ mixed state $\rho_{12}$,
the following inequality holds
\begin{equation}
I_{1}+I_{2}+\left(  {\frac{d}{d-1}\log_{2}d}\right)  \tau_{12}\leq2\log
_{2}d.\label{eq6}%
\end{equation}
\begin{figure}[ptb]
\includegraphics[scale=0.6,angle=0]{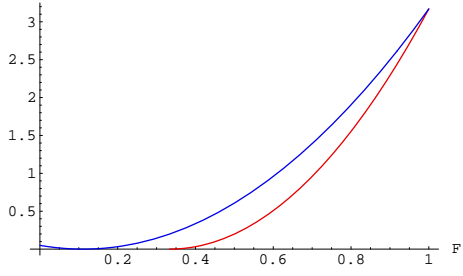}\caption{(color online). Plots of
$I\left(  {\rho_{1}}\right)  +I\left(  {\rho_{2}}\right)  +\left(  {\frac
{d}{d-1}\log_{2}d}\right)  \tau_{12}$ (red line), and $I\left(  {\rho_{12}%
}\right)  $ (blue line) for $d=3$.}%
\label{fig1}%
\end{figure}

Even though this inequality holds for arbitrary $d\times d$ states, for some
special cases we might get a more stringent bound. Consider a class of
isotropic mixed states for $d\times d$ systems. These are states invariant
under $\rho\rightarrow\int{dU\left(  {U\otimes U^{\ast}}\right)  }\rho\left(
{U^{\dag}\otimes U^{\ast\dag}}\right)  $ and can be expressed as
\begin{equation}
\rho_{iso}\left(  F\right)  =\frac{1-F}{d^{2}-1}\left(  {\mathrm{I}-\left\vert
{\Phi^{+}}\right\rangle \left\langle {\Phi^{+}}\right\vert }\right)
+F\left\vert {\Phi^{+}}\right\rangle \left\langle {\Phi^{+}}\right\vert
,\label{eq7}%
\end{equation}
where $\left\vert {\Phi^{+}}\right\rangle =\frac{1}{\sqrt{d}}\sum
\nolimits_{i=1}^{d}{\left\vert {ii}\right\rangle }$ and $F\in\left[
{0,1}\right]  $. For $F\in\left[  {0,\frac{1}{d}}\right]  $ this state is
known to be separable. The tangle for these isotropic states have been
obtained in Ref.~\cite{Rungta:2003}. For $d=3$ the tangle is
\begin{equation}
\tau\left(  {\rho_{12}}\right)  =\left\{  {%
\begin{array}
[c]{l}%
0\ \ \ (F\leq\frac{1}{3})\\
3\left(  {F-\frac{1}{3}}\right)  ^{2}\ \ \ (\frac{1}{3}\leq F\leq1)
\end{array}
}\right.  .\label{eq8}%
\end{equation}

Straightforward calculation shows that the local information for qutrit 1 and
qutrit 2 are both 0. Then the value of $I\left(  {\rho_{1}}\right)  +I\left(
{\rho_{2}}\right)  +\left(  {\frac{3}{2}\log_{2}3}\right)  \tau_{12}$ can be
calculated using Eq.~(\ref{eq8}). On the other hand, we consider the local
information of the composite 2-qutrit system. Using the formula (\ref{eq4}),
we obtain $I\left(  {\rho_{12}}\right)  =\frac{9}{4}\left(  {Tr\rho_{12}%
^{2}-\frac{1}{9}}\right)  =\frac{81F^{2}-18F+1}{32}\log_{2}3$. For a vivid
comparison, we plot $I\left(  {\rho_{1}}\right)  +I\left(  {\rho_{2}}\right)
+\left(  {\frac{3}{2}\log_{2}3}\right)  \tau_{12}$ and $I\left(  {\rho_{12}%
}\right)  $ in Fig.~1. From Fig.~1 we find there always exists the relation
$I\left(  {\rho_{1}}\right)  +I\left(  {\rho_{2}}\right)  +\left(  {\frac
{3}{2}\log_{2}3}\right)  \tau_{12}<I\left(  {\rho_{12}}\right)  $. Based on
this fact, we conjecture that for a general $d\times d$ isotropic mixed state
the following inequality is always true
\begin{align}
& I\left(  {\rho_{1}}\right)  +I\left(  {\rho_{2}}\right)  +\left(  {\frac
{d}{d-1}\log_{2}d}\right)  \tau_{12}\nonumber\\
& \leq\frac{2d^{2}}{d^{2}-1}\log_{2}d\left(  {Tr\rho^{2}-\frac{1}{d^{2}}%
}\right)  .\label{eq9}%
\end{align}

Note that inequality (\ref{eq9}) does not hold for arbitrary $d\times d$
states. It is tempting to define mutual invariant information like the von
Neumann entropy. However, this is not reasonable because there does not exist
the subadditivity relation for invariant information, i.e., $I\left(
{\rho_{12}}\right)  \leq I\left(  {\rho_{1}}\right)  +I\left(  {\rho_{2}%
}\right)  $. It only satisfies the additivity relation like $I\left(
{\rho\otimes\sigma}\right)  =I\left(  \rho\right)  +I\left(  \sigma\right)  $.
In general, we can derive a lower and upper bound of $I\left(  {\rho_{12}%
}\right)  -I\left(  {\rho_{1}}\right)  -I\left(  {\rho_{2}}\right)  $. First,
we can image a $d\times d$ mixed state $\rho_{12}$ is a part of a pure state
$\left\vert \psi\right\rangle _{12R}$, where $R$ denotes a reference system.
Usually the reduced density matrix $\rho_{R}$ also lies in a $d\times d$
dimensional Hilbert space. For the composite $d^{2}\times d^{2}$ pure state
$\left\vert \psi\right\rangle _{12R}$ under the cut $12:R$, we have
\begin{equation}
I\left(  {\rho_{12}}\right)  +I\left(  {\rho_{R}}\right)  +\left(
{\frac{2d^{2}}{d^{2}-1}\log_{2}d}\right)  \tau_{12:R}=4\log_{2}d,\label{eq10}%
\end{equation}
where we have used the complementarity relation for pure states. On the other
hand, we have
\begin{equation}
I\left(  {\rho_{1}}\right)  +I\left(  {\rho_{2}}\right)  +\left(  {\frac
{d}{d-1}\log_{2}d}\right)  \tau_{12}\leq2\log_{2}d.\label{eq11}%
\end{equation}
Combining Eq.~(\ref{eq10}) and Eq.~(\ref{eq11}), we get
\begin{equation}%
\begin{array}
[c]{l}%
I\left(  {\rho_{12}}\right)  -I\left(  {\rho_{1}}\right)  -I\left(  {\rho_{2}%
}\right)  \geq\\
2\log_{2}d-I\left(  {\rho_{R}}\right)  -{\frac{2d^{2}}{d^{2}-1}}\left(
{\log_{2}d}\right)  \tau_{12:R}+{\frac{d}{d-1}}\left(  {\log_{2}d}\right)
\tau_{12},\\
\end{array}
\label{eq12}%
\end{equation}

It is easy to see that the lower bound depends not only on the entanglement of
$\rho_{12}$, but also on the entanglement of $\rho_{12}$ with the reference
system. It should be noted that this bound can be negative value. For example,
consider a seperable two-qubit mixed state ${\rho_{12}}=\frac{5}{12}\left\vert
{00}\right\rangle \left\langle {00}\right\vert +\frac{4}{12}\left\vert
{01}\right\rangle \left\langle {01}\right\vert +\frac{2}{12}\left\vert
{10}\right\rangle \left\langle {10}\right\vert +\frac{1}{12}\left\vert
{11}\right\rangle \left\langle {11}\right\vert $, one can verify that
$I\left(  {\rho_{12}}\right)  -I\left(  {\rho_{1}}\right)  -I\left(  {\rho
_{2}}\right)  =-\frac{5}{54}$. The upper bound of $I\left(  {\rho_{12}%
}\right)  -I\left(  {\rho_{1}}\right)  -I\left(  {\rho_{2}}\right)  $ is
easily obtained. We should only consider a $d\times d$ maximally entangled
pure state $\left\vert \psi\right\rangle _{12}=\frac{1}{\sqrt{d}}%
\sum\nolimits_{j=1}^{d}{\left\vert {ii}\right\rangle }$. The local information
$I\left(  {\rho_{1}}\right)  $ and $I\left(  {\rho_{2}}\right)  $ for this
special pure state are both 0, while $I\left(  {\rho_{12}}\right)  $ reaches a
maximum value $2\log_{2}d$. Therefore we have $I\left(  {\rho_{12}}\right)
-I\left(  {\rho_{1}}\right)  -I\left(  {\rho_{2}}\right)  \leq2\log_{2}d$. The
above analysis suggests that we cannot treat invariant information in the same
way as the von Neumann entropy; sometimes we might resort to the von Neumann
entropy to gain a more comprehensive insight into the quantum information.
Invariant information provides a complement of the von Neumann entropy in the
description of quantum information, but cannot substitute the role of the von
Neumann entropy. Perhaps a deeper investigation of invariant information is desired.

\begin{figure}[ptb]
\includegraphics[scale=0.33,angle=0]{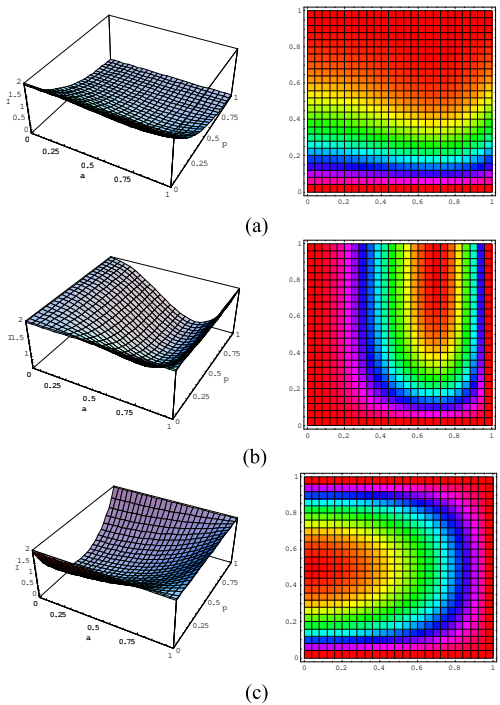}\caption{(color online). The
plots of the local information for the state $\left\vert \psi\right\rangle
_{12}=a\left\vert {00}\right\rangle +\sqrt{1-a^{2}}\left\vert {11}%
\right\rangle $ in the three decoherence processes. The right pannels are
plotted in the density form of the left ones.}%
\label{fig1}%
\end{figure}

So far, we only consider the systems in the isolated environment. One can
easily visualize that the local information will decrease in open systems. In
order to gain more insight into the above question, we analyze the behaviour
of simple two-qubit entangled states under decoherence. We choose the
depolarization, dephasing, and dissipation channel as our toy model for decoherence.

Consider the two-qubit state $\left\vert \psi\right\rangle _{12}=a\left\vert
{00}\right\rangle +\sqrt{1-a^{2}}\left\vert {11}\right\rangle $ where $a$ is
real number with $0\leq a\leq1$. We define $p$ as the degree of decoherence of
an individual qubit, which lies between 0 and 1. Here the value 0 means no
decoherence, and 1 means complete decoherence. The depolarization process with
a decoherence degree $p$ is described by
\begin{equation}
\left\vert i\right\rangle \left\langle j\right\vert \rightarrow\left(
{1-p}\right)  \left\vert i\right\rangle \left\langle j\right\vert
+p\delta_{ij}\frac{I}{2}.\label{eq13}%
\end{equation}
The dephasing process is represented by
\begin{equation}
\left\vert i\right\rangle \left\langle j\right\vert \rightarrow\left(
{1-p}\right)  \left\vert i\right\rangle \left\langle j\right\vert
+p\delta_{ij}\left\vert i\right\rangle \left\langle j\right\vert .\label{eq14}%
\end{equation}
The dissipation is an energy-lossing process, and thus changes the state to a
specific state. We choose $\left\vert 0\right\rangle $ as the ground state.
Then the dissipation process is described by
\begin{equation}%
\begin{array}
[c]{l}%
\left\vert i\right\rangle \left\langle i\right\vert \rightarrow\left(
{1-p}\right)  \left\vert i\right\rangle \left\langle i\right\vert +p\left\vert
0\right\rangle \left\langle 0\right\vert ,\\
\left\vert i\right\rangle \left\langle j\right\vert \rightarrow\left(
{1-p}\right)  ^{\frac{1}{2}}\left\vert i\right\rangle \left\langle
j\right\vert ,\;\;\mbox{where}\;i\neq j.
\end{array}
\label{eq15}%
\end{equation}

After the action of the depolarization channel, the local information of the
2-qubit system is given by
\begin{equation}
\label{eq16}
\begin{array}{l}
 I = \frac{2}{3}\left( {4Tr\rho ^2 - 1} \right) \\
 = \frac{2}{3}\left\{ {4\left[ {a^2\left( {1 - a^2} \right)} \right.}
\right.\left( {2p^4 - 8p^3 + 10p^2 - 4p} \right) \\
 \left. {\left. { + \frac{1}{4}p^4 - p^3 + 2p^2 - 2p + 1} \right] - 1}
\right\} \\
 \end{array}.
\end{equation}
We plot Eq.~(\ref{eq16}) in Fig. 2(a), and the local information under the
dephasing and dissipation processes are plotted in Fig. 2(b) and (c),
respectively. It shows in Fig. 2 that the minimum value of the local
information of the 2-qubit systems is 0 for the depolarization channel while
it is always a positive value for the other two channels. The main reason is
due to the fact that the depolarization channel will transform the initial
state into a totally mixed state after infinite time, while the local
information for maximally mixed state is 0.

Before concluding we would like to stress that our complementarity relation is
based on invariant information. This perspective is different from the methods
used in
Refs.~\cite{Oppenheim:2002,Horodecki:2003,Horodecki:2005,Horodecki1:2005},
where the authors utilized the von Neumann entropy to investigate the
complementarity between local and nonlocal information. The distinct advantage
by using invariant information is based on the following facts. The first is
that invariant information always implies additivity, which is a desired
property to establish complementarity relations. The second is that invariant
information can be directly related to the tangle of $d\times d$ pure states.
The third nice feature of the invariant information is that its definition is
operational. It is obtained by synthesizing the errors of a specially chosen
set of measurements performed on the system. Thus we might get a different
insight into the complementarity relation between the local and nonlocal information.

Summarizing, we have generalized the invariant information introduced by
Brukner and Zeilinger. For $d\times d$ mixed states (where $d$ is prime or the
power of prime), using the invariant information, we establish a
complementarity relation between the local and nonlocal information under the
isolated environment. Furthermore, we show that the nonlocal information has a
direct relation with the entanglement of the system. Some differences between
the invariant information and von Neumann entropy are also discussed. We also
investigate the dynamics of the local information in the open systems through
a simple example.

W.S. thanks Jian-Ming Cai for valuable discussions. This work was
supported by the NNSF of China, the CAS, and the National
Fundamental Research Program (under Grant No. 2006CB921900).

\end{document}